
\input harvmac
\def\half{{1 \over 2}}
\def\dzm{{\partial_z}}

\def\dzp{{\partial _{\bar z}}}

\def\N{{\nabla}}
\def\Nb{{\bar\nabla}}

\def\a {{\alpha}}
\def\b {{\beta}}

\def\ad {{\dot\alpha}}
\def\bd {{\dot\beta}}

\def\t {{\theta}}
\def\ta {{\theta^\alpha}}

\def\tba {{\bar\theta^\ad}}

\def \ad {{\dot \a}}
\def \bd {{\dot \b}}
\def \t {{\theta}}
\def \tb {{\bar\theta}}

\def \Gtp {{\tilde G^+}}
\def \Gtm {{\tilde G^-}}

\def\half{{1 \over 2}}
\def\dzm{{\partial_z}}
\def\dzp{{\partial _{\bar z}}}

\def\a {{\alpha}}
\def\b {{\beta}}

\def\ad {{\dot\alpha}}
\def\bd {{\dot\beta}}

\def\t {{\theta}}
\def\tb {{\bar\theta}}
\def\ta {{\theta^\alpha}}
\def\tba {{\bar\theta^{\ad}}}

\def\N {{\nabla}}
\def\Nb {{\bar\nabla}}
\def\Na {{\nabla_\alpha}}

\Title{\vbox{\hbox{IFUSP-P-1161}}}
{\vbox{\centerline{\bf New Spacetime-Supersymmetric Methods
for the Superstring}}}
\bigskip\centerline{Nathan Berkovits}
\bigskip\centerline{Dept. de F\'{\i}sica Matem\'atica, Univ. de S\~ao Paulo}
\centerline{CP 20516, S\~ao Paulo, SP 01498, BRASIL}
\centerline{and}
\centerline{IMECC, Univ. de Campinas}
\centerline{CP 1170, Campinas, SP 13100, BRASIL}
\bigskip\centerline{e-mail: nberkovi@snfma1.if.usp.br}
\vskip .2in

Talk delivered at
``Strings '95'' Conference at the Univ. of Southern California
\vskip .2in

In this talk, the new spacetime-supersymmetric description of
the superstring is reviewed and some of its applications are described.
These applications include the
manifestly spacetime-supersymmetric calculation of scattering amplitudes,
the construction of a super-Poincar\'e invariant open superstring field theory,
and the beta-function
calculation of low-energy equations of motion in superspace.
Parts of this work have been
done in collaboration with deBoer, van Nieuwenhuizen, Ro\v{c}ek, Sezgin,
Skenderis, Stelle, and especially, Siegel and Vafa.

\Date{June 1995}
\newsec {Introduction}

Ever since the discovery of spacetime-supersymmetry in the superstring,
physicists have looked for a description of the superstring where this
symmetry is manifest. Just as manifest spacetime-supersymmetry simplifies
calculations in supersymmetric particle theories by reducing the
number of Feynmann diagrams, a manifestly spacetime-supersymmetric description
of the superstring removes the need to sum over spin structures, thereby
simplifying the study of multiloop scattering amplitudes. Furthermore, such
a description makes it easier to analyze superstring properties
which depend crucially on the presence of spacetime-supersymmetry, such as
duality and finiteness.

With the exception of the work described here, the only quantizable
description of the superstring with manifest spacetime-supersymmetry is
the light-cone Green-Schwarz description.\ref\GS{M.B. Green and J.H. Schwarz,
Nucl. Phys. B243 (1984) 475\semi
S. Mandelstam, Prog. Theor. Phys. Suppl. 86 (1986) 163.
}\ref\RT{A. Restuccia and
J.G. Taylor, Phys. Rep. 174 (1989) 283.}
This description requires light-cone
gauge fixing of all superstring fields, which breaks manifest
SO($d-1,1$) Lorentz invariance down to SO($d-2$). When calculating scattering
amplitudes using this formalism, non-trivial
interaction-point operators need to be
inserted whereever strings join or split. Since the locations of these
interaction points are complicated functions of the $P^+$ momenta of the
external strings, only four-point tree and one-loop scattering amplitudes
were explicitly calculated using this description (although \RT contains
explicit expressions for multiloop four-point
amplitudes, these expressions contain
unphysical divergences when interaction-points collide).

A similar problem exists for amplitude calculations
using the light-cone Ramond-Neveu-Schwarz
description of the superstring, however it can be overcome by
using the N=1 superconformal invariance of the underlying
covariant RNS description to remove the dependence on the
interaction-point locations.\ref\func{N. Berkovits, Nucl. Phys. B304 (1988)
537.} Unfortunately, a similar procedure
is not possible using the fermionic Siegel invariances
of the standard covariant GS description. However in 1989,
Sorokin, Tkach, Volkov, and Zheltukhin discovered that by introducing
twistor-like variables into the covariant GS action, the fermionic
Siegel invariances could be converted into superconformal invariances.\ref
\STVZ{D.P. Sorokin, V.I. Tkach, D.V. Volkov, and A.A. Zheltukhin,
Phys. Lett. B216 (1989) 302.}
Although the number of classical superconformal invariances for the
$d$-dimensional GS superstring is equal to $d-2$,\ref\tw{N. Berkovits,
Phys. Lett. B241 (1990) 497\semi M. Tonin,
Phys. Lett. B266 (1991) 312\semi E.A. Ivanov and A.A. Kapustnikov,
Phys. Lett. B267 (1991) 175\semi F. Delduc, A. Galperin, P. Howe and
E. Sokatchev, Phys. Rev. D47 (1993) 578\semi
L. Brink, M. Cederwall and C. Preitschopf, Phys. Lett. B311 (1993) 76.}
all but two of the
invariances can be gauge-fixed without ghosts or global moduli. The
remaining invariances form a quantum N=2 superconformal algebra with
$c=3(d-6)/2$, and can be covariantly gauge-fixed with the usual N=2 ghosts
when $d=10$.\ref\hetN{N. Berkovits, Nucl. Phys. B379 (1992) 96.}

In a flat ten-dimensional background, the non-covariant
gauge-fixing of six of the eight
invariances
breaks SO(9,1) Lorentz invariance down to SU(4)$\times$U(1) (this subgroup
can be slightly enlarged to a 25-dimensional subgroup which preserves
pure projective spinors\ref\work{N. Berkovits, work in progress.}).
But for generic compactifications to four
dimensions which preserve
supersymmetry, these extra six invariances are not present so
all of the N=1 SO(3,1) super-Poincar\'e invariance can be made
manifest.\ref\cov{N. Berkovits, Nucl. Phys. B431 (1994) 258.}
For these compactifications, the $c=6$
N=2 superconformal generators
split into a $c=-3$ four-dimensional spacetime
piece, which is independent of the
compactification, and a $c=9$ piece, which is related in
the usual way to a six-dimensional
Calabi-Yau manifold. Also for compactifications
to six dimensions, manifest SO(5,1) super-Poincar\'e invariance can
be preserved and
the generators split into a $c=0$ six-dimensional spacetime piece
and a $c=6$ Calabi-Yau piece.\ref\six{N. Berkovits and C. Vafa,
Nucl. Phys. B433 (1995) 123.}

It is natural to ask what is the relationship between this new N=2
GS description of the superstring and the N=1 RNS description. With a
suitable field redefinition from the N=2 GS matter fields to the
N=1 RNS matter and ghost fields, the N=2 superconformal generators
can be expressed as the stress-tensor, the $b$ ghost, the BRST current,
and the ghost current of the RNS superstring.\ref\equiv{N. Berkovits,
Nucl. Phys. B420 (1994) 332.}  Using the results of
reference \ref\crit{N. Berkovits and C. Vafa, Mod. Phys. Lett. A9
(1994) 653} for critically embedding an N=1 string into an N=2 string,
this shows the equivalence of scattering amplitudes using the two
different descriptions.

In the second section of this paper, the new N=2 description of
the superstring
will be reviewed for the case of four-dimensional compactifications.
In the third section, vertex operators will be constructed and it
will be shown how to calculate manifestly SO(3,1) super-Poincar\'e
invariant scattering amplitudes. Certain ``topological'' multiloop
scattering amplitudes are extremely easy to calculate using these
methods.\six
In the fourth section, a new super-Poincar\'e invariant
open superstring field theory is constructed which
does not suffer from the infinite contact terms
of RNS superstring field theory.\ref\field{N. Berkovits, hep-th 9503099,
to appear in Nucl. Phys. B.} The new superstring
field theory action resembles a WZW action, which can also be
used as a
string field theory action for four-dimensional self-dual Yang-Mills.
In the conclusion, future applications for the new description of the
superstring are discussed, which include coupling to a curved
supergravity/super-Yang-Mills background and using quantum N=2 superconformal
invariance to determine the low-energy equations of motion in
superspace.\ref\stony
{N. Berkovits,J. deBoer, P. van Nieuwenhuizen, M. Ro\v{c}ek, E. Sezgin,
K. Skenderis, K. Stelle, and W. Siegel, work in progress.}

\newsec{The N=2 GS superstring for compactifications to four dimensions}

The worldsheet variables for the four-dimensional part of the N=2
superstring
consist of the spacetime variables, $x^m$ ($m=0$ to 3), the right-moving
fermionic
variables,
$\t^\a$ and $\tb^\ad$ ($\a,\ad=1$ to 2), the conjugate right-moving
fermionic variables, $p_\a$ and $\bar p_\ad$, and one right-moving
bosonic variable, $\rho$.
The chiral boson
$\rho$ is identified with $\rho \pm 2\pi$
and is related
to R-transformations of four-dimensional superspace.
For the heterotic superstring,
one has the usual 32 left-moving fermionic variables, $\chi_I$,
while
for the Type II superstring, one has
the left-moving fermionic variables,
$\t^{*\a}$,$\tb^{*\ad}$,
$p^*_\a$, $\bar p^*_\ad$, and one left-moving
bosonic variable, $\rho^*$.

In conformal gauge, the worldsheet action for the heterotic superstring
variables is:
\eqn\conf{\int d^2 z [\half\dzp x^m \dzm x_m + p_\a \dzp\t^\a +
\bar p_\ad \dzp\tb^\ad -\half\dzp \rho\dzm\rho +\half \chi_I\dzp\chi_I ].}
The free-field OPE's for these worldsheet variables are
\eqn\OPE{x^m(y) x^n(z)\to -\eta^{mn}\log|y-z|,
\quad \rho(y) \rho(z) \to \log(y -z),}
$$p_\a(y)\theta^\b (z)\to {\delta_\a^\b\over{y -z}},\quad
 \bar p_\ad(y)\tb^\bd (z)\to {\delta_\ad^\bd\over{y -z}},\quad
\chi_I(y) \chi_J(z)\to{\delta_{IJ}\over {\bar y -\bar z}}. $$
Note that the chiral boson $\rho$ can not
be fermionized since
$e^{i\rho(y)}~e^{i\rho(z)}~\to e^{2i\rho(z)}(y -z)^{-1}$ while
$e^{i\rho(y)}~e^{-i\rho(z)}~\to (y -z)$.
It has the same behavior as the
negative-energy field $\phi$ that appears when bosonizing the RNS ghosts
$\gamma=\eta e^{i\phi}$ and $\beta=\partial\xi e^{-i\phi}$\ref\FMS{D.
Friedan, E. Martinec and S. Shenker, Nucl. Phys. B271 (1986) 93.}.

These worldsheet GS variables form a representation of an $N=2$
superconformal algebra with $c=-3$. The generators of this algebra are
given by:
\eqn\GSf{L_{d=4}=-\half\dzm x^m \dzm x_m -
p_\a\dzm \t^\a -  \bar p_\ad \dzm\tb^\ad +\half\dzm\rho\dzm\rho}
$$G^+_{d=4}=e^{i\rho} (d)^2 , \quad
G^-_{d=4}=e^{-i\rho} ( \bar d)^2, \quad
J_{d=4}=-i\dzm\rho, $$
where
$$d_\a=p_\a+i\tba\dzm x_{\a\ad}+\half(\tb)^2\dzm\t_\a
-{1\over 4}\t_\a \dzm (\tb)^2,$$
$$ \bar d_\ad= \bar p_\ad
+i\ta\dzm x_{\a\ad}+\half(\t)^2\dzm\tb_\ad
-{1\over 4}\tb_\ad \dzm (\t)^2, $$
and $(d)^2$ means
$\epsilon^{\a\b} d_\a d_\b$. It is straghtforward to check
\ref\siegGS{W. Siegel, Nucl. Phys. B263 (1986) 93.}\
that $d_\a$ and $d^*_\ad$ anticommute with the
$d=4$ spacetime supersymmetries which are generated by
\eqn\GSsusy{
q_\a=\int dz [p_\a -i\tba\dzm x_{\a\ad}-{1\over 4}(\tb)^2\dzm\t_\a],}
$$ \bar q_\ad=\int dz^- [ \bar p_\ad
-i\ta\dzm x_{\a\ad}-{1\over 4}(\t)^2\dzm\tb_\ad],$$
and satisfy the OPE's
\eqn\dO{d_\a(y) V(x,\t,\tb)(z) \to {\N_\a V(z)\over {y-z}},\quad
d^\a(y) \bar d^\ad (z)\to {2i \Pi_z^{\a\ad}\over{y-z}},}
where $V$ is an arbitrary spacetime superfield, $\N_\a=
{\partial\over\partial\t^\a}+i\tb^\ad{\partial\over\partial x^{\a\ad}}$,
and $\Pi_z^{\a\ad}=\dzm x^{\a\ad}-i\t^\a\dzm\tb^\ad
-i\tb^\ad\dzm\t^\a$.

Since the Calabi-Yau manifold is descibed by an N=2 superconformal
field theory with $c=9$,
the combined system
of the four-dimensional GS superstring and the Calabi-Yau manifold
is described by an N=2 superconformal field theory with $c=6$. The
generators of the corresponding N=2 algebra are given by:
\eqn\GSfour{L= L_{d=4}+~L_{CY},\quad
G^-=G_{d=4}^- + ~G^-_{CY}, }
$$G^+=G_{d=4}^+~+ G^+_{CY}, \quad
J=J_{d=4}~+J_{CY},$$
where $[L_{CY},G^-_{CY},G^+_{CY},J_{CY}]$ are the $(N=2,c=9)$
generators describing the Calabi-Yau manifold and
$[L_{d=4},G^-_{d=4},G^+_{d=4},J_{d=4}]$ are the $(N=2,c=-3)$
generators defined in \GSf. Note that integral Calabi-Yau charge is
required for all physical vertex operators since $J=0$ implies that
Calabi-Yau charge is equal to $\rho$ charge, which must be integral in order to
avoid branch cuts with $G^\pm_{d=4}$.

As discussed in \equiv, there is a field transformation from N=2 GS matter
fields into N=1 RNS matter and ghost fields which preserves all
OPE's, maps $G^-$ into
the $b$ ghost, and maps $G^+$ into the RNS BRST current. This
transformation also maps $q_\a$ of \GSsusy
into RNS spacetime-supersymmetry
generators in the $-\half$ picture, and $\bar q_\ad$ into
RNS spacetime-supersymmetry generators in the $+\half$ picture.
Since the generator of $R$-transformations,
\eqn\Rparity{R=\int dz (i\dzm\rho+\half(p_\a \t^\a -\bar p_\ad \tb^{\ad}))}
is mapped into the RNS picture operator,
GS $R$-weight is equal to RNS picture.

\newsec{Scattering Amplitudes}

All physical states of the superstring are represented
by N=2 primary fields, $V$, which are constructed entirely out
of matter fields and are dimension zero (for the heterotic string,
the left-moving part of $V$ is dimension one).
In other words, $L$ and $G^\pm$
have only $(y -z)^{-1}$ singularities with $V$
while $J$ with $V$ has no singularities. The integrated form of the
vertex operators is given by $\int d^2 z |G^- G^+|^2 V$ where
$G^\pm V$ means the
residue of
the single pole in the OPE of $G^\pm$ and $V$.
Under the transformation mapping GS matter fields into
RNS matter and ghost fields, $V$ is mapped into $\xi W$ where $W$ is
an RNS BRST-invariant vertex operator whose picture is equal to its
$R$-weight (i.e., vector bosons are in picture 0, chiral fermions are
in picture $-\half$, and anti-chiral fermions are in picture $+\half$).

For four-dimensional massless superfields in the heterotic string,
the vertex operators take the simple form
$V=  \dzp x^m E_m(x,\t,\tb)$
and $V= (\chi^J f^I_{JK} \chi^K) V_I(x,\t,\tb)$
where $E_m$ is the prepotential
for the supergravity/axion multiplets and $V_I$ is the prepotential for the
Yang-Mills multiplets
(e.g.,
the graviton and axion fields are represented
by $h_{mn} +b_{mn}= \sigma_n^{\a\ad}\N_\a \Nb_\ad E_m$
and the gauge field by $A_{I m}= \sigma_m^{\a\ad}\N_\a \Nb_\ad V_I$
where $\N_\a=\partial_{\t^\a} + i\tb^{\ad} \partial_{\a\ad}$,
$\Nb_\ad=\partial_{\tb^{\ad}} + i\t^{\a} \partial_{\a\ad}$).
The condition of being primary implies the on-shell conditions
\eqn\onshell
{\N^2 E_m
=\Nb ^2 E_m =\eta^{mn}\partial_m E_n=\eta^{mn}\partial_m\partial_n E_p
=\N^2 V_I=\Nb ^2 V_I=
\eta^{mn}\partial_m\partial_n V=0.}
The integrated form of the above vertex operators is obtained
by hitting $V$ with $G^+$ and $G^-$ and, for massless superfields, takes the
form:
\eqn\integrated{U=
\int d^2 z ~
(\bar d^\ad ~\N^2\Nb_\ad  +d^a ~\Nb^2\N_\a}
$$ +\half(
\dzm\tba~ \Nb_\ad +\dzm\ta~\Na + \Pi^{\a\ad}~(\Na\Nb_\ad-\Nb_\ad\Na) ))^2 V.
$$

A similar structure exists for
four-dimensional massless fields in the Type II string, which
are
represented by the N=(2,2) primary field
$V=E(x, \t,\tb,\t^*,\tb^*)$ (e.g., the graviphoton field strength is
given by
$$F_{mn}= \sigma_{mn}^{\a\b}\Nb^2 \N_\a(\Nb^*)^2\N^*_\b  E
+ \sigma_{mn}^{\ad\bd} (\N^2)\Nb_\ad(\N^*)^2\Nb^*_\bd E.$$
Note that in both the heterotic and Type II strings, the axion and
graviton vertex operators are constructed from the
same prepotential. This fact will be relevant when
constructing non-linear sigma models for these strings
in a curved background.

One way to calculate scattering amplitudes is to introduce N=2 ghosts,
construct an N=2 BRST operator and picture-changing operators, and integrate
correlation functions of BRST-invariant vertex operators on N=2 super-Riemann
surfaces. However, a simpler way is to twist the N=2 string (which allows
the central charge of the matter fields to cancel) and use N=4 topological
techniques developed in \six. The advantage of this topological method is that
there is no need to introduce N=2 ghosts or to integrate over N=2 super-moduli.
Since the N=2 GS matter system is equivalent to the N=1 RNS matter and ghost
systems, it should not be surprising that there is no need to introduce a new
set of ghosts.

The N=4 topological method involves twisting the $c=6$ stress-tensor $L$ to
$L+\half J$, and defining two new bosonic and fermionic generators:
\eqn\new{J^{++}=e^{\int^z J},\quad
J^{--}=e^{-\int^z J}, \quad\Gtp=G^- J^{++},\quad \Gtm=G^+ J^{--}}
where $G^\mp J^{\pm\pm}$ means the residue of the single pole in their OPE.
Using the generators defined in \GSfour,
\eqn\newtwo{J^{++}=e^{-i\rho+\int^z J_{CY}},\quad
J^{--}=e^{i\rho-\int^z J_{CY}},}
$$\Gtp=e^{-2i\rho+\int^z J_{CY}} \bar d^2
+ G^-_{CY}
e^{-i\rho+\int^z J_{CY}}, \quad\Gtm=e^{2i\rho-\int^z J_{CY}} d^2
+ G^+_{CY}
e^{i\rho-\int^z J_{CY}}.$$
It is easy to check that when $c=6$, these generators combine with the
original N=2 generators of \GSfour
to form a small N=4 algebra. Note that under
the GS $\to$ RNS field transformation, $J^{++}=c\eta$, $J^{--}=b\xi$,
$\Gtp=\eta$, and $\Gtm=bZ$ where $Z$ is the N=1 picture-changing operator.

The condition that $V$ is primary implies the equation of motion
$\Gtp G^+ V=0$ where multiplication by $G^+$ or $\Gtp$ signifies
taking its contour integral (after twisting, $G^+$ and $\Gtp$ are
spin-one). Gauge transformations of $V$ are
$\delta V=G^+ \Lambda+\Gtp \bar\Lambda$, which leave
$\Gtp G^+ V$ invariant since $G^+$ and $\Gtp$ anti-commute with each other.
Since under the GS $\to$ RNS field transformation, $\Gtp V=W$ is the
RNS vertex operator, $\Gtp=\eta$, and $G^+$ is the RNS BRST operator,
these conditions reproduce the standard cohomology of physical superstring
states. Note however that there are no square-root cuts in the OPE's
of \OPE, so there is no need to perform a GSO projection when using the
GS variables.

The prescription for R-invariant
tree-level scattering amplitudes is given by the
$N$-point correlation function
\eqn\tree{A_0=
<G^+ V_1(z_1) ~ \Gtp V_2(z_2) V_3(z_3) U_4 ... U_N >,}
and for R-invariant $g$-loop amplitudes is given by
\eqn\ampfour{A_g=
 \prod_{j=1}^{3g-3}\int d^2 m_j
\prod_{i=1}^g  {\tilde G}^+(v_i)
 \prod_{j=1}^{2g-2} (\mu_j \Gtm)
 \prod_{j={2g-1}}^{3g-2} (\mu_j G^-) (\mu_{3g-3} J^{--})
U_1 ... U_{2g}>,}
where $\mu_j$ are the Beltrami differentials and $U_i=\int d^2 z G^- G^+ V$
(scattering amplitudes which violate R-invariance by an amount $m$ correspond
to calculations on N=2 surfaces of U(1) instanton number $m$, and therefore
require $m$ $G^+$'s to be converted to $\Gtp$'s in the above formulas).
It is easy to relate these amplitudes to the standard RNS prescription by
using the transformation that maps the GS matter fields into the RNS
matter and ghost fields. Under this transformation, \tree and \ampfour get
mapped into
$$A_0=
<W_1(z_1) ~Z W_2(z_2)~ \xi W_3(z_3) U_4 ... U_N >,$$
$$A_g=\prod_{j=1}^{3g-3}\int d^2 m_j
\prod_{i=1}^g
<\prod_{i=1}^{g} \eta (v_i)
 \prod_{j=1}^{2g-2} (\mu_j b Z)
 \prod_{j={2g-1}}^{3g-2} (\mu_j b) (\mu_{3g-3} b\xi)
U_1 ... U_{2g}>,$$
where $Z$ is the RNS picture-changing operator and $W_i=\xi V_i$ are the
RNS vertex operators
(for R-invariant amplitudes, the sum of the pictures
of $W_i$ is zero). Since this RNS
calculation is in the ``large'' Hilbert space involving the zero mode
of $\xi$, one has to insert $g$ zero modes for $\eta$.
As was shown in reference \ref\Carow{U. Carow-Watamura,
Z. Ezawa, K. Harada, A. Tezuka and S. Watamura, Phys. Lett. B227
(1989) 73.}, the locations of these insertions are
determined by
restrictions on the $\phi$ loop-momentum, and the combined
$(\xi,\eta)$ and $\phi$ correlation functions reproduce the usual
Verlinde prescription for the $(\beta,\gamma)$ ghost correlation
function.\ref\Ver{E. Verlinde and H. Verlinde, Phys. Lett. B192 (1987) 95.}
Note that without these restrictions on the
$\phi$ loop-momentum, the amplitude is divergent since $\phi$ is a
negative energy field (i.e. its kinetic term appears with the wrong sign).

Since $\rho$ is also a negative-energy field,
similar restrictions must be imposed on its
loop-momentum in \ampfour. As in the RNS case,
the locations of the
$\Gtp$ insertions are determined by these
restrictions on the $\rho$ momentum, and
work is in progress on using this fact to obtain
explicit expressions for arbitrary
multiloop GS amplitudes (note that \tree contains no such subtleties
and therefore provides the first manifestly
super-Poincar\'e invariant formula for tree-level superstring amplitudes).

For multiloop scattering amplitudes which depend in a trivial way
on the $\rho$ variable, these subtleties can be ignored
and explicit expressions can be obtained. For example,
for the scattering of $2g-2$ chiral graviphotons,
these expressions reproduce the ``topological'' amplitudes which
were originally obtained in a more complicated way
using the RNS formalism.\ref\ant{I. Antoniadis, E. Gava, K.S. Narain and
T.R. Taylor, Nucl. Phys. B413 (1994) 162.}
Similar GS techniques have also been used
in six dimensions to obtain explicit topological expressions for the
scattering of $4g-4$ chiral graviphotons.

\newsec{Open superstring field theory}

As was discussed in section 3, the linearized equations of motion
and gauge invariances for the N=2 vertex operators are
\eqn\lin{\Gtp G^+ V=0,\quad \delta V=G^+ \Lambda +\Gtp\bar\Lambda,}
rather than the usual $QW=0$, $\delta W=Q\Lambda$.
This suggests that their non-linear generalizations in superstring
field theory may look different from the Chern-Simons generalizations
$QW=W^2$, $\delta W= Q \Lambda - [W ,\Lambda]$.\ref\wit{E. Witten,
Nucl. Phys. B268 (1986) 253\semi E. Witten, Nucl. Phys. B276 (1986) 291.}
This is good news since all known Chern-Simons versions of superstring
field theory require midpoint operator insertions which lead to either
contact terms with infinite coefficients or unphysical solutions which need
to be removed by hand.\ref\contact{C. Wendt, Nucl. Phys. B314 (1989) 209
\semi C.R. Preitschopf, C.B. Thorn and S. Yost, Nucl. Phys. B337 (1990)
363\semi I. Ya. Aref'eva and P.B. Medvedev, Phys. Lett. B202 (1988) 510.}

By comparing with the tree-level scattering amplitude of \tree, one can show
that the correct non-linear generalization of \lin for
four-dimensional open string superfields,
$$V(x^m(\sigma),\t^a(\sigma),
\tb^\ad(\sigma), p^\a (\sigma),\bar p^\ad (\sigma),
\partial_\sigma \rho(\sigma)),$$ is
\eqn\nonlin{\Gtp (e^{-V} G^+ e^V)=0,\quad
\delta V=(G^+ \Lambda) e^V +e^V(\Gtp\bar\Lambda),}
where all string superfields (including those coming from a Taylor series
expansion of the exponential) are multiplied using Witten's
half-string overlap.\wit Since the contributing pieces from
$G^+$ and $\Gtp$ are $e^{i\rho} d^2$ and
$e^{-2i\rho+J_{CY}} \bar d^2$, the gauge invariances for the
massless component of $V$ (which only depends on $x$, $\t$, and $\tb$)
are those of the super-Yang-Mills prepotential,
$$\delta V=(D^2 \lambda)e^V + e^V (\bar D^2\bar\lambda),$$
where  $\Lambda=e^{-i\rho}\lambda$,
$\bar\Lambda=e^{2i\rho+J_{CY}}\bar\lambda$,
and $\lambda,\bar\lambda$ are unconstrained spacetime superfields.
Note that all matter string superfields are required to be U(1)
neutral, so the gauge parameters carry U(1) charge $-1$.

The action which yields \nonlin as the equation of motion resembles
the two-dimensional Wess-Zumino-Witten action where the derivatives
$\dzm$ and $\dzp$ are replaced by the operators $G^+$ and $\Gtp$.
This action, which includes only the contribution of the four-dimensional
superfields, is
\eqn\sa
{Tr [\,\, (e^{-V}G^+ e^V)(e^{-V}\Gtp e^V)-\int_0^1 dt
(e^{-tV}\partial_t e^{tV})
\{ e^{-tV} G^+ e^{tV},
e^{-tV}\Gtp e^{tV} \}] }
where the trace over string
states is defined as in Witten's open string field
theory.\wit

With a slight modification, it is also possible to include the contribution
of superfields which depend on the compactification manifold. The complete
SO(3,1) super-Poincar\'e invariant superstring field theory action for
arbitrary supersymmetric compactifications can be found in reference \field.
Although this action contains terms of all orders in the string field,
it differs from the RNS action in that
the coefficients of all such terms are explicit and finite.

Since the action in \sa only requires the existence of a $c=6$
N=2 superconformal field theory, it can also be used as a string field
theory action for four-dimensional super-Yang-Mills. For the N=2 string
which represents self-dual Yang-Mills, the worldsheet variables are
$x^i$ ($i$=1 to 2), $\bar x_i$, $\psi^i$, $\bar\psi_i$,
and the relevant superconformal generators are $G^+=\bar\psi_i\dzm x^i$
and $\Gtp=\epsilon^{ij}\bar\psi_i\dzm \bar x_i$ (after twisting,
$\psi^i$ is spin-one and $\bar\psi_i$ is spin-zero). If
$\Phi(x^i(\sigma),\bar x_i (\sigma),\psi^i(\sigma),\bar\psi_i(\sigma))$
is the string field, then $\Gtp(e^{-\Phi}G^+ e^{\Phi})=0$ implies that
its massless mode $\phi$ (which only depends on $x$ and $\bar x$)
satisfies $\partial^i(e^{-\phi}\partial_i e^{\phi})=0$, which is the
equation of motion for the Yang scalar of self-dual Yang-Mills.\ref\oogur
{H. Ooguri and C. Vafa, Nucl. Phys. B367 (1991) 83.}

\newsec{Future Applications}

In this talk, a new description of the superstring was used to calculate
manifestly spacetime-supersymmetric scattering amplitudes and to
construct a super-Poincar\'e invariant superstring field theory action
which does not suffer from the contact-term problems of the RNS action.
There are various further applications for this new description.

One obvious application is to couple the worldsheet GS variables to
a curved supergravity/super-Yang-Mills background, and to use
beta-function techniques to obtain the low-energy superstring
equations of motion
in superspace.\ref\stony
{N. Berkovits,J. deBoer, P. van Nieuwenhuizen, M. Ro\v{c}ek, E. Sezgin,
K. Skenderis, K. Stelle, and W. Siegel, work in progress.}
This is done by generalizing
the heterotic worldsheet action of \conf to:
\eqn\hetconf{{1\over{\alpha'}}\int d^2 z [\half \Pi^a_z \Pi_{\bar z \,a}
+ d_\a \Pi_z^\a +
\bar d_\ad \Pi_z^\ad +B_{AB} \Pi^A_z \Pi^B_{\bar z}
-\half\dzp \rho\dzm\rho +\half \chi_I D_{\bar z} \chi_I }
$$+\phi (r+if) +\bar\phi (r-if) +\psi (e^{i\rho} d^\a \N_\a \phi)+
 +\bar\psi (e^{-i\rho} \bar d^\ad \Nb_\ad \bar\phi)]$$
where $A$ are flat-space indices, $M$ are curved space indices,
$\Pi_z^A=E_M^A \dzm Z^M$,
$\Pi_{\bar z}^A=E_M^A \dzp Z^M$, $E_M^A$ is the supergravity vielbein, and
$B_{AB}$ is an anti-symmetric two-form whose field strength is
$H_{ABC}=\nabla_{[A} B_{BC]}$. In a flat background where
$B_\a^m=E_\a^m=i\sigma_{\a \ad}^m \tb^\ad$, it is straightforward to check
that this action reduces to \conf.

However, in order to be classically superconformally invariant, the field
strength $H_{ABC}$ must be set equal to the torsion $T_{ABC}$ (this can
be seen at the level of vertex operators from the fact that $g_{mn}$ and
$b_{mn}$ are components of the same prepotential $E_m$). So $B_{MN}$ is not
an independent superfield and one still needs to find a coupling for the
dilaton/axion multiplet.\ref\sigma{W. Siegel, Phys. Lett. B211 (1988)
55\semi
S.J. Gates Jr., P. Majumdar. R.N. Oerter and A.E. van de Ven,
Phys. Lett. B214 (1988) 26}

A similar situation arises in the bosonic string where the dilaton couples
to the two-dimensional curvature, so it is natural to try to couple this
dilaton/axion multiplet to the N=(2,0) supercurvature. This can be done in
an N=(2,0) super-reparameterization invariant manner by adding the term
\eqn\dil{\int d^2 z d\kappa R \Phi +
\int d^2 z d\bar\kappa \bar R\bar\Phi=}
$$\int d^2 z [\Phi (r+if) +\bar\Phi (r-if) +\psi (e^{i\rho} d^\a \N_\a \Phi)+
 +\bar\psi (e^{-i\rho} \bar d^\ad \Nb_\a \bar\Phi)]$$
where $\kappa$
and $\bar\kappa$ are the worldsheet anti-commuting parameters,
$R=\psi+(r+if)\kappa$ and
$\bar R=\bar\psi+(r-if)\bar\kappa$ are the worldsheet chiral and
anti-chiral superfields describing N=(2,0) supercurvature ($r$ is the
ordinary curvature, $f$ is the U(1) field strength, and $\psi,\bar\psi$
are the gravitino field strengths), $\Phi(x,\t)$ and $\bar\Phi(x,\tb)$ are
spacetime chiral and anti-chiral superfields for the dilaton multiplet
(the dilaton is $(\Phi+\bar\Phi)|_{\t=\tb=0}$ and the axion is
$i(\Phi-\bar\Phi)|_{\t=\tb=0})$.\ref\mysigma{N. Berkovits,
Phys. Lett B304 (1993) 249.}

The term $\int d^2 z d\kappa R\Phi$ makes sense since $G^-$ has no singularity
with $\Phi$, and therefore $\Phi$ is worldsheet chiral as well as spacetime
chiral. Note that the dilaton zero modes couples to the Euler number of
the surface and the axion zero mode couples to the U(1) instanton number.
So just as the string coupling constant can be absorbed into the dilaton field,
the string theta-parameter (which counts the violation of R-invariance in
scattering amplitudes) can be absorbed into the axion field.

As in the bosonic string, the supercurvature term \dil is not classically
superconformally invariant, but because it is higher order in $\alpha'$,
its classical variation is expected to cancel the quantum variation of
\hetconf. Work is in progress on checking that the sum of \hetconf and
\dil is superconformally invariant at the quantum level when the
low-energy equations of motion are imposed
on the supergravity/super-Yang-Mills background.\stony
Note that the compactification-dependent fields have been frozen in this
sigma model, so the only effect of the compactification is to
increase the central charge by $+9$.

Other possible applications of this new description include the
coupling of Type II strings in N=2 d=4 supergravity backgrounds\ref\sieg
{N. Berkovits and W. Siegel, work in progress}, comparison of multiloop
scattering amplitudes and effective actions
for different six-dimensional superstrings\ref\oog{N. Berkovits,
H. Ooguri, and C. Vafa, work in progress},
and the construction of a super-Poincar\'e invariant version of
closed superstring field theory.

This work has been financially supported by the
Conselho Nacional de Pesquisa.

\listrefs
\end